\documentclass[slac_one]{revtex4}
\usepackage{graphicx}
\usepackage{fancyhdr}
\usepackage{atlasphysics}
\newcommand{\mt}{$M_{top}$}
\newcommand{\ttb}{\mbox{$t\bar{t}$}}

\begin{document}
\title{Top quark mass measurement with ATLAS} 

%

\author{Antoine Marzin (on behalf of the ATLAS collaboration)}
\affiliation{CEA-Saclay, IRFU, SPP, F-91191 Gif-sur-Yvette, France}

\begin{abstract}

The top quark mass measurement with ATLAS in the lepton ($\ell$ = e,$\mu$) plus jets channel is summarized
from the perspective of the first fb$^{-1}$ of data taking.  
Using the invariant mass of the three jets arising from the hadronic side as the estimator of the top quark mass,
a precision of the order of 1 to 3.5 \gev\ on the top quark mass measurement should be achievable with 1 \ifb\ of collected data, assuming a jet energy scale uncertainty of 1 to 5 $\%$.

\end{abstract}

\maketitle

\thispagestyle{fancy}


\section{Introduction} 

With a mass close to the scale of the electroweak symmetry breaking, the top quark mass \mt\ is a particularly important parameter
of the Standard Model (SM). For instance, we can test the consistency of any theory (SM and its extensions) and probe Physics beyond the SM 
 by comparing the measurements of the electroweak precision observables with the theoretical predictions of the model under consideration.
As a distinctive feature, the top quark mass enters quadratically in the $W$~boson mass $M_W$ while the dependence on the Higgs boson mass $M_H$ 
is only logarithmic. Consequently, accurate measurements of both $M_W$ and \mt\ are required to provide stringent theoretical constraints on $M_H$
in order to test the consistency of the SM in the hypothesis of a SM Higgs boson discovery.

At the LHC, top quarks will be produced mainly in $t \bar{t}$ pair through the hard process $gg~\rightarrow~{\ttb}$~(90$\%$) and $q\bar{q}~\rightarrow~{\ttb}$~(10$\%$) with 
a cross section at the next to leading order of 833 pb. Therefore, about 800 000 $t \bar{t}$ pairs will be produced with the
integrated luminosity of 1~$fb^{-1}$ expected in 2009. This huge statistic available at the LHC will allow us to perform
a direct top quark mass measurement by determining the peak of the invariant mass of the particles arising from the top quark decay.    
With a width $\Gamma$ $\simeq$ 1.5 \gev, roughly an order of magnitude 
greater that $\Lambda_{QCD}$, the top quark has such a short lifetime that it decays before 
hadronizing and almost exclusively through the mode $ t \, \rightarrow \, W b$. I summarize here the measurement of \mt\ in the lepton ($\ell$ = e,$\mu$) plus jets channel,
where one of the $W$ bosons decays hadronically while the other one decays leptonically (the analysis is detailed in \cite{CSC}). This channel is the most
promising one since it constitutes the best compromise between a high branching ratio    
and a high signal over background ratio, the lepton of the final state allowing a good rejection of the QCD background.

\section{Event selection and background}

The major source of physical background (summarized in Table 1) are single top events and $W$+jets production with 
leptonic $W$~boson decay. An analysis performed on fast simulation has shown that backgrounds from QCD and $b \bar{b}$
production are negligible after the leptonic cuts (isolated lepton, $p_T$, \met).

%
\begin{table}
\begin{center}
\caption{\label{selection} Number of remaining events for the main backgrounds to the semi-leptonic signal 
after the successive cuts.}
\begin{tabular}{||l|c|c|c|c||} 
\hline
Process & Number of events & 1 isolated lepton 	&  $\geq$ 4 jets    & 2 b-jets         \\ 
	& ($\mathcal{L}$~=~1~fb$^{-1}$)& \pt\ $\geq$ 20 \gev  & \pt\ $\geq$ 40 \gev & \pt\ $\geq$ 40 \gev \\ 
        &           & and \met $\geq$ 20\gev & &\\  \hline
\hline \hline
Signal $\ell$ = e,$\mu$  +jets &  313200 & 132380 & 43370 & 15780\\
\hline \hline
$W$~boson backgrounds  &  9.5 $\times 10^5$ & 154100 &9450 & 200 \\ \hline \hline
all-jets (top pairs)& 466 480 &  1 020 & 560 & 160  \\ \hline 
di-lepton (top pairs)& 52 500 &  16 470 & 2 050 & 720 \\ \hline \hline
single top: t, $W$t and s channels& 91 810 &  33 470 & 2 011 & 505  \\ \hline 
\end{tabular}
\end{center}
\end{table}

All particles forming the \ttb\ final state are required to lie within $|\eta| \leq 2.5$.
In order to reduce the QCD background and the \ttb\ fully hadronic and di-lepton processes, we require \met\ higher than 20 \gev\ and exactly one
isolated lepton ($\ell$~=~e,~$\mu$) with $p_T \geq 20 \gev$. To reduce the $W$+jets background,  exactly 2 b-tagged jets
are required with $p_T \geq  40 \gev$. Finally, we require at least two light jets with $p_T \geq  40 \gev$. This cut leads to an important
loss of efficiency since only 34 $\%$ of W bosons with hadronic decay have both jets passing this requirement, but jets below
40~\gev\ are known to be less precisely calibrated. Moreover, 30 $\%$ of signal events have more than two light jets because of initial and
final state radiation (ISR/FSR); therefore, we require at least two light jets (and not exactly two).  

The efficiency of the selection with respect to $l$+jets events is 5 $\%$ and, starting from a  signal over background ratio very unfavorable, the successive cuts 
allow us to reach a signal over background ratio of 10 at this step of the reconstruction.

\section{Hadronic W boson mass reconstruction}

\begin{figure}[htbp]
 \begin{minipage}[t]{.45\linewidth}
\centering
\includegraphics[height=5cm,width=8cm]{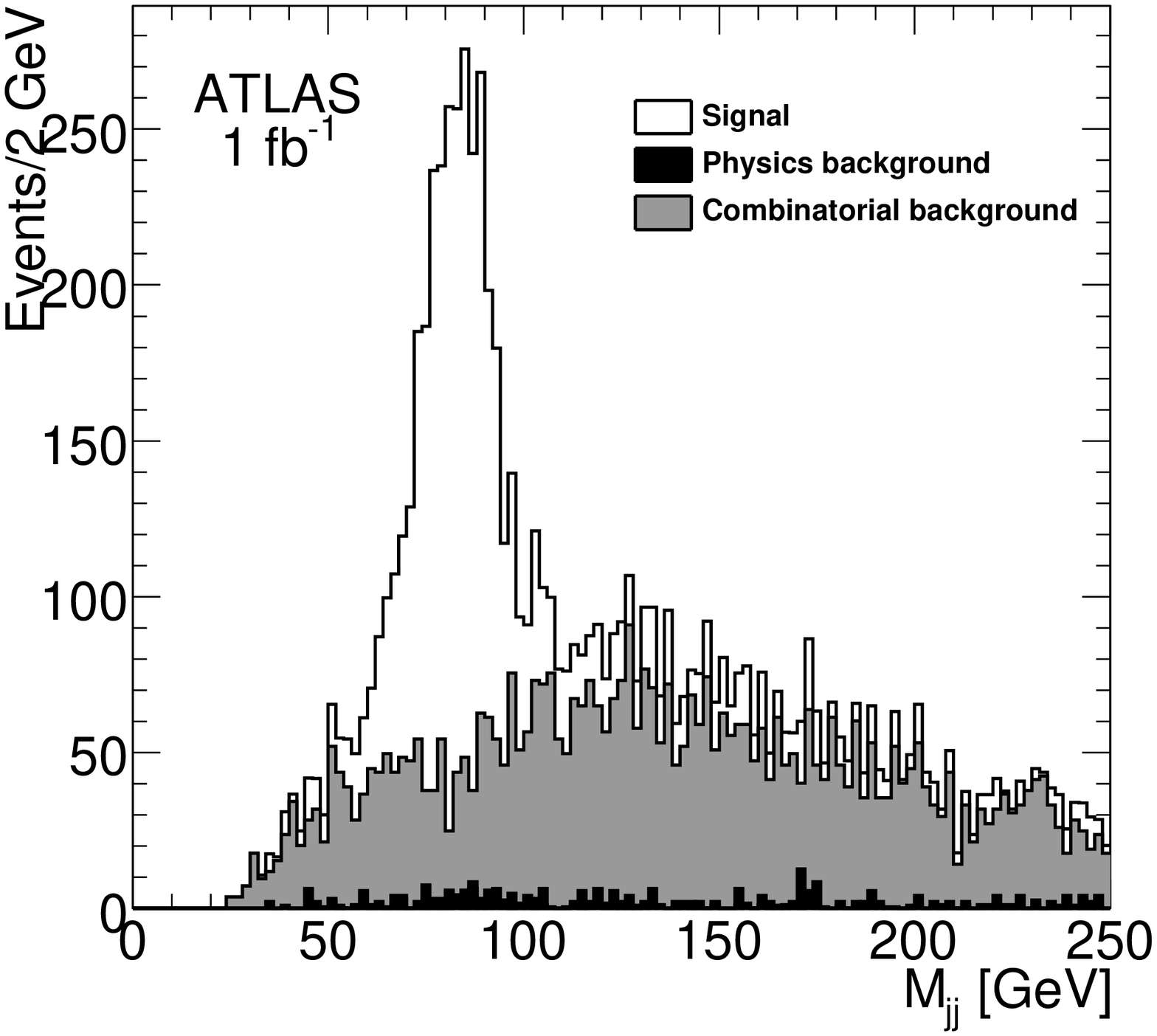}
\caption{\label{dijet} Invariant mass of light jet pairs for events  with only two light jets.}
\end{minipage} \hfill  
\begin{minipage}[t]{.45\linewidth}
\centering
\includegraphics[height=5cm,width=8cm]{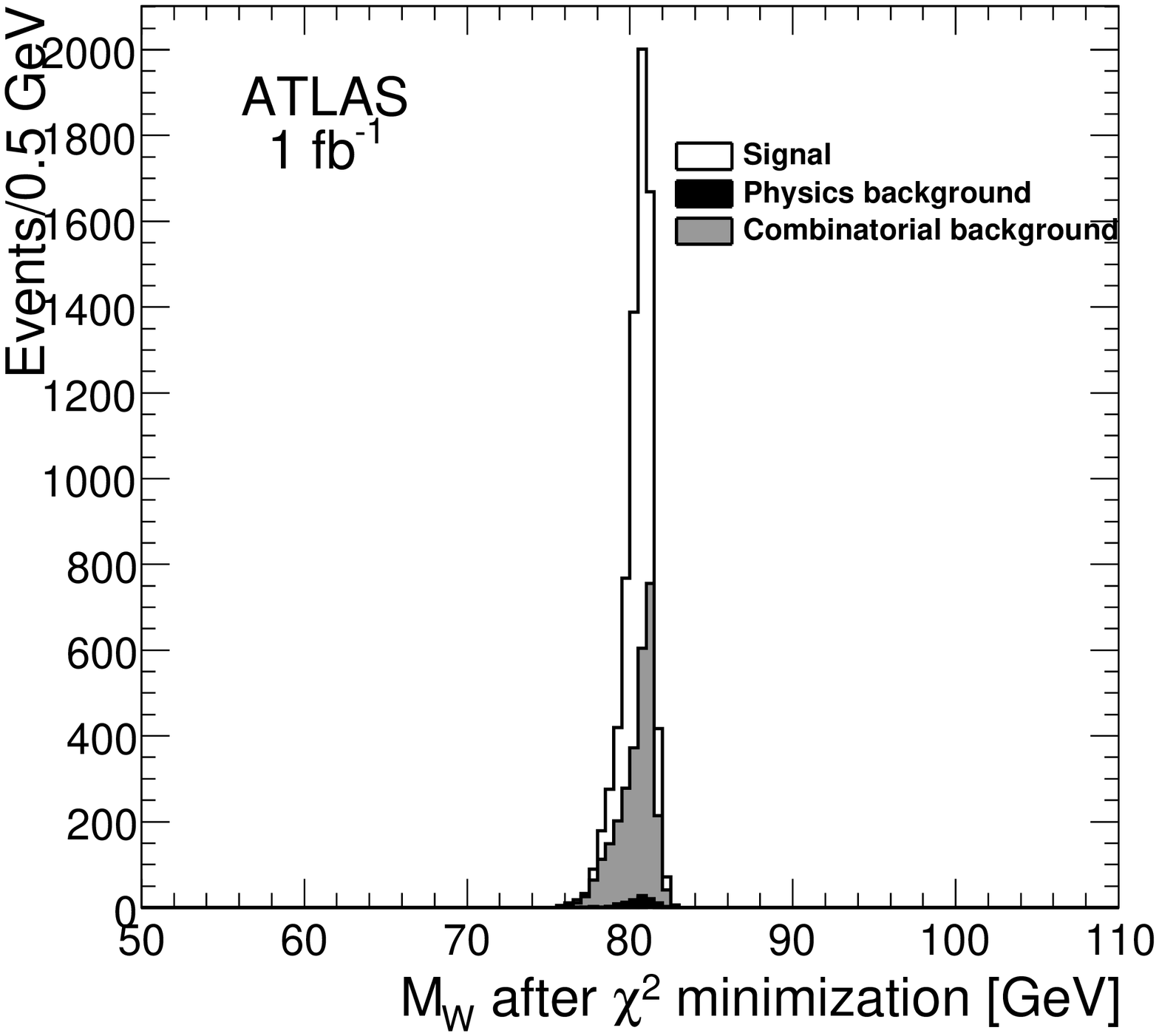}
\caption{Hadronic $W$~boson mass after the $\chi^{2}$ minimization.}
\end{minipage} \hfill 
\end{figure}

For the remaining events with more than two light jets in the final state, we have to find the two jets arising from the hadronically-decaying
$W$ boson. The hadronic $W$ boson candidates are selected in a mass window of $\pm$ 30 \gev\ around the peak value of the dijets distribution
of events reconstructed with exactly two light jets (Fig. \ref{dijet}). We then perform an event by event $\chi^2$ minimization for each possible way of pairing two light jets
and choose the pair with the smallest $\chi^2$ as the hadronic $W$ boson candidate. The $\chi^2$ formula contains a term
which constrains the invariant mass of the two jets $M_{\mathrm{jj}}$ to the $W$ boson mass and width from the Particle Data Group as follows:
$$
\chi^2  =  \frac{(M_{\mathrm{jj}}(\alpha_1,\alpha_2) -  M_{\mathrm{W}}^{\mathrm{PDG}})^2}{(\Gamma_{\mathrm{W}}^{\mathrm{PDG}})^2}  +  \frac{(E_{\mathrm{j1}}(1 - \alpha_1))^2}{\sigma_{E_1}^2}  +  \frac{(E_{\mathrm{j2}}(1 -  \alpha_2))^2}{\sigma_{E_2}^2}.
$$
The main asset of this method is to reduce the systematic uncertainty due to the light jet energy scale since we perform
an event by event in-situ rescaling of light jets by constraining the jet energies to their measured value within 
the jet energy resolution $\sigma_E$.
Further on, only events with a hadronic $W$ boson candidate
within a mass window of $\pm$~2~$\Gamma_{\mathrm{M_W}}^{\mathrm{PDG}}$ ($\Gamma_{\mathrm{M_W}}^{\mathrm{PDG}}$ = 2.1 \gev) are kept.
\section{Top quark mass determination with the hadronic side}

Once the hadronic $W$ boson candidate is selected, we have to choose among the two b-jets the right one to be associated
with the hadronic $W$ boson in order to reconstruct the top quark mass. Severals methods have been investigated and we obtain
a better purity by picking the b-jet which is the closest to the hadronic $W$ boson.

The final efficiency of the reconstruction  with respect to $l$+jets events is 2.22~$\pm$~0.03~$\%$ with a top purity of 40.2~$\pm$~0.8~$\%$.
The reconstructed top quark mass distribution is shown on Fig. 3; the fit is the sum of a Gaussian for the signal and a 
third degree polynomial for the combinatorial background.  
For a generated mass of 175~\gev, we reconstruct \mt~=~175.0~$\pm$~0.2~\gev\ with a width equal to 11.6~$\pm$~0.2~\gev.
\begin{figure}[htbp] \begin{minipage}[h]{.45\linewidth}
\centering
\includegraphics[height=4.8cm,width=8cm]{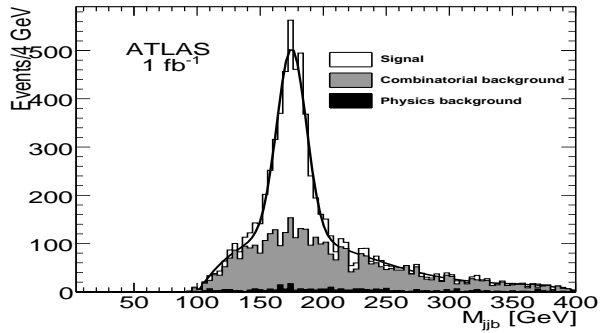}
\caption{The hadronic top quark mass, fit with a Gaussian and a third order polynomial, scaled to 1 fb$^{-1}$. }
\end{minipage}  \hfill 
\begin{minipage}[h]{.45\linewidth}
\begin{center}
\vspace{0.4cm}
\begin{tabular}{||l|c||}\hline
 Systematic uncertainty & Shift on \mt \\
\hline \hline
Light jet energy scale & 0.2 \GeV{}/$\%$  \\ \hline
 b jet energy scale & 0.7 \GeV{}/$\%$ \\ \hline
 ISR/FSR & $\simeq$ 0.3 \GeV{} \\ \hline
 b quark fragmentation & $\leq$ 0.1 \GeV{} \\ \hline
 Background  & negligible \\ \hline
Method & 0.1 to 0.2 \gev \\ \hline
 \hline
\end{tabular}
\end{center}
\vspace{0.65cm}
\begin{flushleft}
Table II: Systematic uncertainties on the top quark mass measurement in the semi-leptonic channel.
\end{flushleft}
\end{minipage}  \hfill
\end{figure} 

\section{Systematic uncertainties}

The statistical uncertainty on \mt\ will be quickly negligible at the LHC; with 1 few fb$^{-1}$ of integrated luminosity, the experimental error on \mt\
will be dominated by the systematic uncertainties summarized in Table 2. Since \mt\ is measured as the invariant mass of three jets, the jet energy scale
constitutes the main source of systematic uncertainties. Its incidence on the top quark mass measurement is estimated
by multiplying separately the light jet and b-jet momenta by a rescaling factor without modifying \met. The resulting top quark mass
being linearly dependent on the rescaling factor, we can express the related uncertainty as a shift on \mt\ by percent of jet energy scale
miscalibration. Note the reduced dependence on the light jet energy scale compared to the b-jet energy scale thanks to the $W$ mass
constraint in the $\chi^2$ minimization. 

\section{Conclusion and perspectives}

The statistical uncertainty being negligible at the LHC, the error on the top quark mass measurement will be dominated by the systematic uncertainties 
and a precision of the order of 1 to 3.5 \gev\ on \mt\ should be achievable with 1 \ifb\ of collected data, assuming a jet energy scale uncertainty of 1 to 5 $\%$.
Several studies have been performed using a relaxed requirement on the b-tagging to increase the efficiency  during the commissioning phase of
data taking. Even with one or no b-tagged jet, we still have a good signal over background ratio and, using a kinematic fit
on the entire final state in order to decrease the combinatorial background, we should reach a precision
on \mt\ of about 5 \gev\ for 100 pb$^{-1}$ assuming a jet energy scale uncertainty of 5 $\%$. The development of other methods for the top quark mass
measurement with ATLAS are in progress in the three different channels: the matrix element method, the template method and the b decay length method which exploits the
correlation between \mt\ and the b quark momentum when the top quark decays at rest.

\end{document}